\documentclass[aps,prb,twocolumn,showpacs,10pt,superscriptaddress]{revtex4-1}
\usepackage{graphicx,amsmath,amssymb,amsfonts,sidecap,color}

\usepackage{ dsfont }

\usepackage{color}

\newcommand{\be}{\begin{equation}}
\newcommand{\ee}{\end{equation}}

\newcommand{\mysection}[1]{{\it #1.}}

\begin{document}

\title{Topological Superconductivity and Majorana Fermions in RKKY Systems}

 \author{Jelena Klinovaja}
 \affiliation{Department of Physics, University of Basel,
              Klingelbergstrasse 82, CH-4056 Basel, Switzerland}
               \author{Peter Stano}
 \affiliation{Department of Physics, University of Basel,
              Klingelbergstrasse 82, CH-4056 Basel, Switzerland}
 \affiliation{Institute of Physics, Slovak Academy of Sciences, 
 845 11 Bratislava, Slovakia}
                \author{Ali Yazdani}
\affiliation{Joseph Henry Laboratories and Department of Physics,
Princeton University, Princeton, New Jersey 08544}
               \author{Daniel Loss}
 \affiliation{Department of Physics, University of Basel,
              Klingelbergstrasse 82, CH-4056 Basel, Switzerland}

\date{\today}


\begin{abstract}
We consider quasi one-dimensional RKKY systems in proximity to an $s$-wave superconductor.
We show that a  $2k_F$-peak in the spin susceptibility of the superconductor  in the one-dimensional limit supports helical order of localized magnetic moments via RKKY interaction,
where $k_F$ is the Fermi wavevector. 
The magnetic helix is equivalent to a uniform magnetic field and very strong spin-orbit interaction (SOI) with an effective SOI length $1/2k_F$.
We find the conditions to establish such a magnetic state in  atomic chains and semiconducting nanowires with magnetic atoms or nuclear spins. 
Generically, these systems are in a topological phase with Majorana fermions. The inherent self-tuning of the helix to $2k_F$ eliminates the need to tune the chemical potential.
\end{abstract}

\maketitle

\mysection{Introduction}
Majorana Fermions (MFs) \cite{Majorana}
have attracted wide attention due to their exotic non-Abelian statistics  and their promise for topological quantum computing, \cite{Kitaev,Alicea_2012} also fueled by recent experiments 
searching for MFs. \cite{Ando,Mourik,Rokhinson,Goldhaber,Heiblum,Xu,Marcus}
The crucial ingredient for most MF proposals are helical spin textures leading to an exotic $p$-wave pairing  due to proximity effect with an ordinary $s$-wave supercondcutor. \cite{hasan_review, TI_review,Alicea_2012,MF_Sato,Majorana_PRL}
The associated helical modes, which transport opposite spins in opposite directions, are proposed to exist in various systems. \cite{kane_graphene,hasan_review,streda,nanotube_helical,Bilayer_MF,nanoribbon_KL,MOS_KL, TI_review,sato_splitter} A well-known mechanism responsible for helical modes is spin-orbit interaction (SOI) of Rashba type. \cite{Alicea_2012,MF_Sato,Majorana_PRL} 
However, quite often, an additional uniform magnetic field is needed, and the spin polarization of the helical modes is not ideal but depends on the SOI strength  \cite{streda}.  While intrinsic values of SOI are limited by material parameters, the recently proposed  synthetic SOI produced by a helical magnetic field can reach extraordinary values that are limited only by the spatial period of the helical field $2\pi/k_n$. \cite{Braunecker_Loss_Klin_2009} Such helical fields can be engineered with nanomagnets, \cite{Braunecker_Loss_Klin_2009,Two_field_Klinovaja_Stano_Loss_2012, nanoribbon_KL,MOS_KL,Flensberg_magnets,exp_magnets} or, more atomistically, can emerge from  helical spin chains due to  anisotropic exchange and Dzyaloshinskii-Moryia interaction. \cite{atoms_exp,atoms_theory_Princeton,Nagaosa,Akhmerov}
The equivalence between spectra of wires with intrinsic and with synthetic SOI has opened up new platforms for helical modes and MFs.
\cite{Braunecker_Loss_Klin_2009,Flensberg_magnets, Two_field_Klinovaja_Stano_Loss_2012, nanoribbon_KL,MOS_KL} However, in all these setups the chemical potential must be tuned inside the gap opened by the magnetic field so that the Fermi wavevector $k_F$ is close to $k_n/2$. 
This poses additional challenges on experimental realizations by requiring  wires with a high tunability of the Fermi level and high mobility down to ultra-low densities. 

Thus, it is natural to ask if helical modes exist in low-dimensional superconductors such that the system automatically tunes itself to $k_n=2k_F$. 
Surprisingly, the answer turns out to be affirmative for a rather broad class of  systems. These are RKKY systems that consist of localized magnetic moments coupled by itinerant electrons via the
Rudermann-Kittel-Kasuya-Yosida (RKKY) interaction. \cite{RKKY_R,RKKY_K,RKKY_Y} It was recently discovered that in such systems in the quasi one-dimensional (1D)  limit  the moments form 
 a spin helix  leading to a Peierls-like gap at the intrinsic Fermi level such that $2k_F=k_n$. \cite{braunecker2009:PRB,Braunecker_Loss_Klin_2009,dominik,meng}
However, these results were obtained for semiconducting or metallic systems in the normal phase, relying on the presence of gapless itinerant electrons to transmit the RKKY interaction. Thus,  it is not clear that the same ordering mechanism can also develop in the superconducting regime where the spectrum of the electrons is gapped. In the present Letter we address this issue in detail and demonstrate that the helical order arising from RKKY interaction survives also in 1D superconductors. As a prototype for our model we consider  atomic chains \cite{atoms_exp} and semiconducting nanowires with magnetic atoms or nuclear spins placed on top of a bulk $s$-wave superconductor. 
We show that these setups are generically deep inside the topological phase and host MFs without requiring any fine-tuning at all.

\begin{figure}[!b]
 \includegraphics[width=\columnwidth]{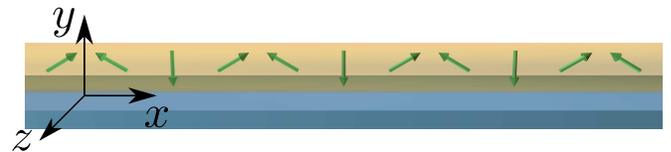}\\
 \caption{Sketch of  a wire (yellow cylinder) aligned along the $x$-axis  brought into contact with an $s$-wave superconductor (blue slab). Localized magnetic moments (green arrows) are ordered into a spin helix by RKKY interaction transmitted by electrons. }
 \label{model}
\end{figure}

\mysection{Model} We consider a quasi 1D superconducting wire aligned along the $x$-axis with embedded localized magnetic moments, see Fig. \ref{model}. The quasi-one-dimensionality means that only the lowest subband of the wire is occupied, \cite{footnote1} and the wavefunction in the transverse direction is given by $\psi(y,z)\equiv \psi(\bf r_\perp)$.
The 1D $s$-wave superconductor is described by the Hamiltonian
\be
H=\int {\rm d}x\, \left[\psi^\dagger \mathcal{H}_0(x) \psi + \Delta_s \left(\psi_\uparrow \psi_\downarrow + {\rm h.c.} \right) \right],
\label{eq:Htot}
\ee
where $\psi(x)=(\psi_\uparrow(x),\psi_\downarrow(x))$ with $\psi_\sigma (x)$ being an annihilation operator acting on an electron at position $x$ with spin $\sigma$. The superconducting coupling parameter $\Delta_s \geq 0$ arises 
from the proximity effect. Here,
$\mathcal{H}_0(x) = -\hbar^2\partial_x^2/2m -\mu_F$,
where $m$ is the effective mass and $\hbar \hat k = -i \hbar \partial_x$ the momentum operator. The energy  is taken from the Fermi level, $\epsilon_k = \hbar^2 (k^2-k_F^2) / 2 m$, where the Fermi wavevector $k_F = \sqrt{2m\mu_F}/\hbar$ is set by the chemical potential $\mu_F$. 
The quasiparticle energy in the superconductor is given by $\eta_k = \sqrt{ \epsilon_k^2 +\Delta_s^2}$. 

The interaction between itinerant electron spins and localized magnetic moments ${\tilde {\bf I}}_i$ at position ${\bf R}_i=(x_i, {\bf r}_{\perp,i})$ is described by 
the  Hamiltonian density
\be
\mathcal{H}_{int}(x) = \frac{\beta}{2} \sum_i  {\tilde {\bf I}}_i \cdot \boldsymbol{\sigma} |\psi({\bf r}_{\perp,i})|^2 \delta(x-x_i),
\label{eq:Hx}
\ee
with the coupling strength $\beta$ being a material constant. Here, $\boldsymbol{\sigma}$ is the vector of Pauli matrices acting in electron spin space.
In the following we solve the interacting Hamiltonian $H+H_{int}$ on a mean field level: We first integrate out the superconducting condensate in leading order $\beta$ to derive an
effective RKKY interaction 
 for the subsystem of localized moments. We find its groundstate and quantify conditions under which it is stable. Assuming these conditions are fulfilled, we derive an effective Hamiltonian for the electron subsystem.

\mysection{RKKY in a 1D superconductor}
The localized moments spin-polarize the conducting medium, which influences other moments and results in the RKKY interaction.  We now introduce an effective 1D model of magnetic moments, with a notation suitable for magnetically doped semiconductor wires, and extend it later to other realizations. 
We approximate the transverse wavefunction profile $|\psi({\bf r}_\perp)|^2$ by a constant, the inverse of the cross-section area $A$. We assume there are $N_\perp$ localized moments on a transverse plane. Though $N_\perp$ might be large, once the magnetic order is established, these spins are collinear, since the spin excitations within the locked transverse plane are energetically much more costly than excitations we will consider below and can  therefore be neglected. \cite{braunecker2009:PRB} Thus, a transverse plane is  assigned a single (effective) spin of length $I=N_\perp {\tilde I}$. Neighboring planes are separated by a distance of the order of the lattice constant $a$, and the density of moments is parametrized as $N_\perp = \alpha \rho_0 A a$, with $\alpha$ being the fraction of cations replaced by magnetic atoms and $\rho_0=4/a^3$ the cation density in zinc-blende materials. 

\begin{figure}
\includegraphics[width=0.4\textwidth]{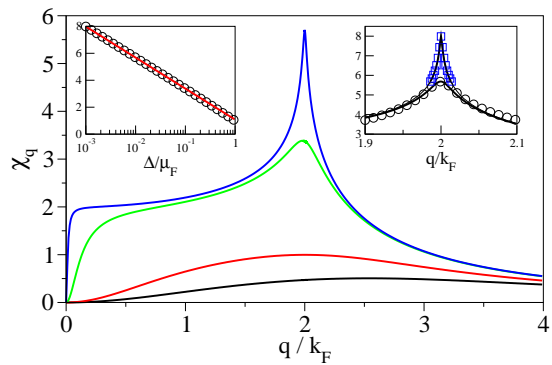}
\caption{(Color online) One-dimensional superconductor susceptibility divided by $k_F/8\pi a\mu_F$. Main: Susceptibility given in Eq.~\eqref{eq:chiq} for $\Delta_s/\mu_F = 2$, 1, 0.1, and 0.01, {\it resp.} (curves from bottom up). Insets: comparison of exact [symbols; Eq.~\eqref{eq:chiq}] and interpolation [solid line; Eq.~\eqref{eq:chiqinterp}] values. Left: at $q=2k_F$. Right: for $\Delta_s/\mu_F = 0.01$ (circles), and 0.001 (squares).}
\label{fig:chiq} 
\end{figure}

With these definitions, the  RKKY interaction between localized moments becomes 
\be
H^{RKKY} = - \sum_{i,j} J_{ij}{\bf I}_i \cdot {\bf I}_j,
\label{eq:RKKY}
\ee
where the long range coupling 
$J_{ij} = -\frac{2\beta^2}{A^2}  \chi(x_i-x_j)$ is given by the  static spin susceptibility $\chi$ of the 1D superconductor.
In Fourier space, $\chi_q = (1/a) \int {\rm d}x \, \exp(i q x) \chi(x)$,
the  susceptibility  at zero temperature  is given by \cite{abrikosov}
\be
\chi_q = -\frac{1}{4 a L}\sum_k \frac{\eta_k \eta_{k+q}-\epsilon_k \epsilon_{k+q}-\Delta_s^2}{\eta_k \eta_{k+q}}\frac{1}{\eta_k + \eta_{k+q}}.
\label{eq:chiq}
\ee
Here, $L=N a$ is the system length. The finite temperature corrections to $\chi$,  being exponentially small if $\Delta_s \gg k_B T$, are neglected. For a plot of $\chi_q$, see Fig.~\ref{fig:chiq}.

Though $\chi_{q=0}=0$ and thus the total magnetization is strictly zero, \cite{anderson1959:PR}
the superconductor responds at non-zero momenta $q$. If $\Delta_s\leq \mu_F$, the susceptibility develops a peak at $q=2k_F$, which gets more pronounced as the ratio $\Delta_s/\mu_F$ drops. Examining the limiting cases analytically, we suggest the following interpolation formula 
\be
\chi_q \approx -\frac{k_F}{8\pi a \mu_F} \ln \left( \frac{e^{1/2+\gamma}}{\sqrt{\Delta_s^2/\mu_F^2 + (1-q/2k_F)^2/2}} \right),
\label{eq:chiqinterp}
\ee
in the regime of $\Delta_s\lesssim\mu_F$ and $|q-2k_F|\ll k_F$. Here, $\gamma\approx0.58$ is the Euler gamma constant. Comparing to numerics, we find that it is an excellent approximation for the peak height at $q=2k_F$, while away from this point it differs from the exact result by a broad non-resonant background only (see insets of Fig.~\ref{fig:chiq}).

\begin{table}
\begin{tabular}{cccc}
\hline\hline
& chain & \quad magnetic wire \qquad & \quad nuclear wire \qquad\\
\hline
material \qquad & Fe & GaMnAs & InAs\\
$a$ &  0.3 nm 
& 0.565 nm & 0.605 nm\\
$g_e$ & 2 & -0.44 & -8\\
$\tilde I \cdot g_s \cdot \mu$ & $2 \cdot 2 \cdot \mu_B$ & $5/2 \cdot 2 \cdot \mu_B$ & $9/2 \cdot 1.2 \cdot \mu_N$\\
$\beta$ & 1.6 meV nm$^3$
& 9 meV nm$^3$ & 4.7 $\mu$eV nm$^3$\\
$\alpha$ &1 & 0.02 & 1\\
$\mu_F$ &10 meV & 20 meV & 1 meV\\
$\Delta_s$ & 1 meV & 0.5 meV & 0.1 meV\\
\hline
$T_c$ &  $14$ K & 2 K  & 7.4 mK\\
$B_c$ &$5$ T & 0.7 T & 0.4 T\\
$\Delta_{m}$ & 6 meV & 5 meV & 0.2 meV\\
$\xi$ & 4 nm & 0.4 $\mu$m& 0.5 $\mu$m\\
\hline\hline
\end{tabular}
\caption{Chosen parameter values: material, lattice constant $a$, 
electron $g$-factor $g_e$, magnetic moment $\mu_S=\tilde I g_s\mu$, exchange coupling $\beta$, moment doping $\alpha$, Fermi energy $\mu_F$, superconducting gap $\Delta_s$, and the resulting scales: critical temperature $T_c$, critical field $B_c$, effective helical field strength $\Delta_{m}$, and MF localization length $\xi$ for different realizations: magnetic atom chain, magnetically doped nanowire, and nanowire with nuclear spins. The effective mass for ${\rm GaMnAs}$ (InAs) is $m= 0.067m_e$ ($0.027 m_e$).}
\label{tab:params}
\end{table}

{\it Magnons.}
The classical ground state of Eq.~\eqref{eq:RKKY} with $J$ peaked at finite momentum is a helically ordered pattern, ${\bf I}_i = \mathcal{R}_{ {\bf h}, 2k_F x_i} [ \boldsymbol{\mathcal{I}} ]$ with $\mathcal{R}$ a 3x3 matrix corresponding to a vector rotation around the helical axis ${\bf h}$ and angle $2 k_F x_i$, and $\boldsymbol{\mathcal{I}} \equiv {\bf I}_1 \perp {\bf h}$ is the spin at the wire end. We find the excitations of the corresponding quantum system (magnons) by Holstein-Primakoff representation of the effective spins by bosons. \cite{holstein1940:PR} Magnons, labelled by momentum $k$, have energies
\be
\hbar \omega_k = \frac{\tilde I}{\sqrt{2}} \sqrt{\left(2J^{||}_{2k_F}-J^{||}_{2k_F+k}-J^{||}_{2k_F-k}\right) \left( J^{||}_{2k_F}-J^{\perp}_{k} \right) },
\label{eq:magnons}
\ee
where we introduced the upper index on the tensor $J$ for its value in the helical plane ($||$) and in a direction perpendicular to it ($\perp$). We consider first the isotropic case, $J_q^{||}=J_q^\perp$. The spectrum is gapless, as the two terms under the square root in Eq.~\eqref{eq:magnons} become zero at $k=0$ and $k=2k_F$, respectively. Close to these values the dispersion is linear in $k$, as the function $J$ has a maximum at $2k_F$. 
The helical order critical temperature
 follows as (for details see App. \ref{Magnons})
 \be
 k_B T_c \sim I^2 J_{2k_F} = -\langle H^{RKKY} \rangle/N \equiv E_h,
 \label{eq:Tc}
 \ee
 so that it is given by the RKKY energy scale with expectation value taken in the helical ground state. 
 The critical length turns out to be extremely long and of no concern (see App. \ref{Magnons}).

\begin{figure*}[!tb]
 \includegraphics[width=2 \columnwidth]{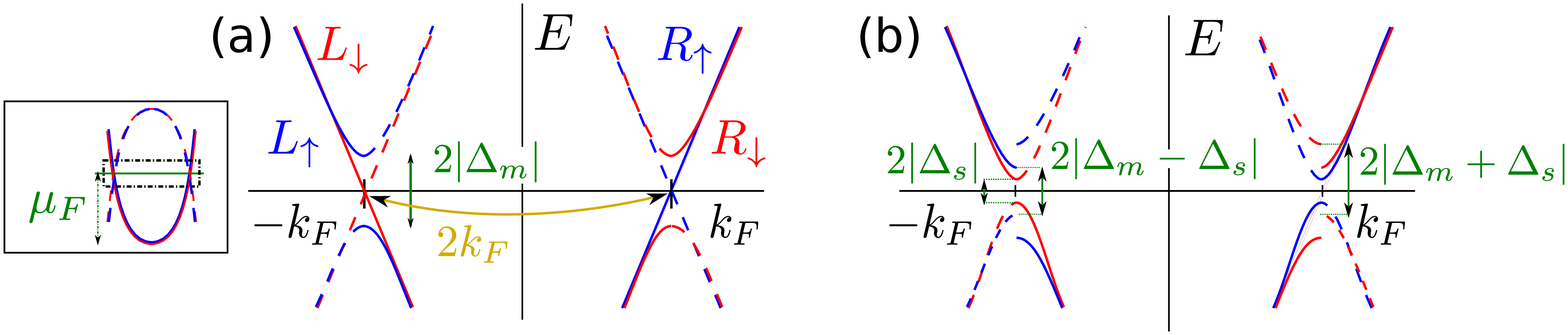}\\
 \caption{(a) Electron (solid lines) and hole (dashed lines) energy spectrum around the Fermi points $\pm k_F$ of a wire in the presence of a helical magnetic field described by $\mathcal H_0 + \mathcal H_m$. The inset shows the entire spectrum of $\mathcal H_0$ with  $\mu_F$ being counted from the band bottom. 
 The interaction term $\mathcal H_{m}$ leads to a resonant scattering between right movers with spin down $R_\downarrow$ and left movers with spin up $L_\uparrow$ and results in a partial gap $2\Delta_m$ in the spectrum. Away from the Fermi points spin up (blue lines) and spin down (red lines) states stay degenerate.   
 (b) The superconducting pairing term of strength $\Delta_s$, which couples electron-hole states with opposite spins and momenta, gaps all branches of the spectrum [see Eqs. (\ref{E_1})-(\ref{E_2})] except for the special case  $\Delta_s=\Delta_m$, which marks the transition between trivial and topological phases.}
 \label{spectrum}
\end{figure*}

\mysection{Realizations}
Next, we discuss three  realizations of our system: magnetically doped semiconductor nanowire, such as GaMnAs, clean III-V semiconductor nanowire with high nuclear spin isotopes, such as InAs, and a chain of magnetic atoms on a superconductor surface, such as Fe on Nb. The parameters and resulting scales are summarized in Tab.~\ref{tab:params}. We include 
the effective field $\Delta_{m} = \alpha \beta \rho_0 \tilde I/2$ the  electrons feel once the order is established, and
$B_c = \min\{E_h/\mu_S ,\Delta_m/g_e\mu_B \}$ gives a rough estimate of the critical field $B_c$ destroying the helical order.

For a semiconductor nanowire with a moderate doping of 2\% we obtain a critical temperature of Kelvins and stability with respect to magnetic fields in the range of Tesla. Such nanowires can be grown very clean, with mean free paths of order microns if undoped, and are  tunable by gates. Magnetic dopants will unavoidably induce some disorder. This problem is absent in the case of an undoped wire with nuclear spins, though there the critical temperature becomes very small, 7 mK, because of the weak hyperfine interaction. We still include this realization for comparison, as establishing helical order in a nuclear wire was considered for electrons in a Luttinger liquid regime. \cite{dominik,braunecker2009:PRB} Like there,
 we expect interactions (neglected here) to enhance $T_c$ by a factor of 2-4.
We took InAs as the nuclear wire material and, to simplify notation in Tab.~\ref{tab:params}, we neglected the contribution of As nuclei, thus $\alpha=1$.

An exciting possibility is offered by magnetic monoatomic chains which can be fabricated on metal \cite{atoms_exp} or superconducting surfaces using either growth
or atomic manipulation techniques with the STM. \cite{atoms_theory_Princeton,yazdani} For the latter system, the formulas we give are valid upon putting $N_\perp=1$, $\alpha=1$, $A=a^2$  with $a$ being the interatomic distance.
As an estimate, we take $\beta /a^3=6$~meV and a hopping matrix element $t=10$ meV.  \cite{atoms_exp,atoms_theory_Princeton,yazdani} The stability of the helical order is rather high, comparable to the stability of the underlying superconductor, so that possible differences in these parameters are not relevant for us. 
What is important here is that  the atomistic exchange and possible spin-orbit based Dzyaloshinskii-Moriya interaction should be negligible for the RKKY-induced magnetism. Under this condition, the magnetic helix pinned to the Fermi level is to be expected.

{\it Majorana Fermions without tuning.}  From now on we consider a helical order of localized magnetic moments to be established. These moments $\tilde{\bf I}_i$ are aligned along the polarization vector ${\bf n} (x)=\cos (2k_F x ) \hat x+  \sin (2k_F x) \hat y$,
that is perpendicular to the helix-axis ${\bf h} \equiv \hat z$. 
Such an ordered state acts back on the electrons via 
$\mathcal{H}_{int}(x)$ [see Eq.~\eqref{eq:Hx}]
that takes the form of an effective Zeeman term
\begin{equation}
\mathcal H_{m}=  \Delta_{m} {\bf n}(x)\cdot {\boldsymbol \sigma},
\end{equation}
where $\Delta_{m} = \alpha \beta \rho_0 \tilde I /2$ is the strength of the helical field assumed to be positive from now on.

The energy spectrum of a system described by $\mathcal{H}+\mathcal H_{m}$ can be easily found by rewritting all electron operators in terms of slow-varying right- ($R_\sigma$) and left- ($L_\sigma$) movers: $\psi_\sigma(x) = R_\sigma(x) e^{ik_Fx} + L_\sigma(x) e^{-ik_Fx}$.
\cite{MF_wavefunction_klinovaja_2012} The helical field  with a pitch  $\pi/k_F$ results in a resonant scattering between right movers with spin down and left movers with spin up, see Fig. \ref{spectrum}a. Taking this resonance into account, we arrive at the effective Hamiltonian
$\tilde {H} =\frac{1}{2}\int dx\ \phi^\dagger(x)  \tilde {\mathcal{H}} \phi(x)$ with the Hamiltonian density
\begin{align} 
\tilde {\mathcal{H}} = -i \hbar \upsilon_F  \tau_3\partial_x  + \frac{\Delta_m}{2}  \eta_3 (\sigma_1\tau_1+\sigma_2\tau_2) + \Delta_s  \eta_2\sigma_2\tau_1,
\label{static}
\end{align}
where the Pauli matrix $\tau_i$ ($\eta_i$) acts in  left-right mover (electron-hole) space, and $\phi=(R_\uparrow, L_\uparrow, R_\downarrow, L_\downarrow, R^\dagger_\uparrow,  L^\dagger_\uparrow, R^\dagger_\downarrow, L^\dagger_\downarrow)$. The Fermi velocity is given by $\hbar \upsilon_F = (\partial \epsilon_k/\partial k)|_{k=k_F}$. The diagonalization of $\tilde {\mathcal{H}} $ gives us the bulk  spectrum,
\begin{align}
&E_\pm^{(1)} = \pm \sqrt{(\hbar \upsilon_F k)^2 + \Delta_s^2},\label{E_1} \\
&E_\pm^{(2,\pm)} = \pm \sqrt{(\hbar \upsilon_F k)^2 + (\Delta_m\pm \Delta_s)^2}. \label{E_2}
\end{align}
Here, $k$ is the momentum eigenvalue 
defined close to the Fermi points $\pm k_F$. The Hamiltonian $\mathcal{H}$ belongs to the topological class $\rm BDI$ \cite{Ryu}, so the system can potentially host MFs. The transition between a topological phase (with MFs) and a trivial phase (no MFs) is related to closing and reopening of an energy gap. In our case, the system is gapless if $\Delta_m=\Delta_s$, see Eq. (\ref{E_2}) for $E_\pm^{(2,-)}$. Straighforward calculations \cite{MF_wavefunction_klinovaja_2012, Two_field_Klinovaja_Stano_Loss_2012} lead us to
the  topological criterion $\Delta_m>\Delta_s$. If it is satisfied, the system is in the topological phase.
We note that for all three realizations considered above $\Delta_m\gg\Delta_s$ (see Tab. \ref{tab:params}), so the system is automatically deeply in the topological phase without any need for parameter tuning.

The MF localization length $\xi$ is determined by the smallest gap in the system, $\xi =  \hbar \upsilon_F/\Delta_s$. If the distance $L$ between MFs localized at opposite ends of the wire  is smaller than  $\xi$, these two MFs combine into an ordinary fermion of non-zero energy. \cite{numerics_Diego}  The numerical values for $\xi$ listed in Tab.~\ref{tab:params} are well below a micrometer. In addition, the coupling between MFs can take place via the bulk superconductor. \cite{sc_gap} However, this channel is also efficiently suppressed in long wires.

We have  tested our model numerically. As expected, the presence of MFs in the topological phase is stable against fluctuations of hopping parameters, the superconductivity strength, and the local chemical potential. We believe that disorder effects, which challenge an observation and identification of MFs, \cite{numerics_Diego,disorder_berlin,disorder_aguado,disorder_potter,disorder_altland,disorder_beenakker,disorder_sarma} can be efficiently suppressed in our setup. Unlike to Rashba nanowires, where the charge density is limited by the (usually weak) Rashba SOI, we can work at much higher densities benefitting from charge impurities being screened.
However, if $\mu_F$ is increased, the critical temperature $T_c$ goes slowly down [as $1/k_F$, see Eq. (\ref{eq:Tc})].  For an atom chain, which can be charged by gates, this is irrelevant as $T_c$ is very high to begin with.
For a semiconducting nanowire, the decrease of $T_c$ can be prevented by increasing magnetic doping.

\mysection{Conclusions}
We have introduced a new class of superconducting systems based on RKKY interactions which feature magnetic helices with a pitch given by half of the  Fermi wavelength
in the quasi one-dimensional limit. As a result, the superconductor becomes topological and hosts MFs without the need to tune the chemical potential. We have proposed candidate systems such as  chains of magnetic atoms
and semiconducting nanowires with nuclear spins or magnetic dopants. 

\acknowledgments
We acknowledge support by the SNF, NCCR QSIT, COQI-APVV-0646-10,
NSF-DMR1104612, and ARO.

\appendix

\section{Spin susceptibility}

The spin susceptibility, the expression for which we give in  the main text, see Eq.~(4), is defined by the relation 
\be
\langle {\bf s}(x) \rangle = \int {\rm d}x^\prime \chi(x-x^\prime) 
\mu {\bf B} (x^\prime),
\tag{S1}
\ee
for the expectation value of the spin density ${\bf s}(x)=\psi^\dagger(x)(\boldsymbol{\sigma}/2)\psi(x)$ arising in response to a magnetic field perturbation $H_m = -\int {\rm d}x\,\mu{\bf B}(x) \cdot {\bf s}(x)$.

\section{Magnons \label{Magnons}}

First we address the question of  the critical temperature and length associated with fluctuations  in low dimensions.
Magnons are excitations of the helical ground state of localized magnetic moments. Without magnons the order is perfect---the total spin polarization $P$ is maximal---while each magnon diminishes it by one unit of $\mu_B$. Magnons have Bose-Einstein statistics, and at finite temperature $T$ we get
\be
P(T)=NI-\sum_{k \neq 0, 2k_F} \frac{1}{\exp(\hbar \omega_k/k_B T)-1},
\label{eq:polarization}
\tag{S2}
\ee
with the magnon dispersion $\omega_k$ given in Eq.~(6) in the main text.
The two terms with exactly zero energy were excluded from the summation, as these do not diminish the polarization, but rather transform one ground state into another. Namely, if $J$ is isotropic, the ground state has continuos symmetry (the full spin rotational symmetry), which, in our model, translates into the symmetry of rotation of $\boldsymbol{\mathcal{I}}$ (spin polarization vector at the wire end) around ${\bf h}$ (the vector defining the helical plane) and the rotation of ${\bf h}$ itself. As hinted by Eq.~(6) of the main text, these two transformations correspond to the two Goldstone bosons $k=0$, and $2k_F$, respectively. In the limit $L\to \infty$ the sum in Eq.~\eqref{eq:polarization} is converted into an integral which necessarily diverges because of vanishingly small energies of magnons in the vicinity of the Goldstone bosons.

From the previous discussion it follows that the order can be established only if the system symmetries are broken, for example, by an anisotropy in the RKKY tensor, $J^{||}\neq J^\perp$. 
Such an anisotropy  pins  the helical axis to certain relative direction to the wire axis, wire-substrate interface, and crystallographic axes. We expect the difference $J^{||}-J^\perp$ to be of the order of $J^{||}$ itself, resulting in the gap $\hbar \omega_{2k_F} \sim S J_{2k_F}$. Alternatively, an easy axis $(K<0)$ or easy plane $(K>0)$ anisotropy seen by the localized spin $H_i=K ({\bf S}_i \cdot {\bf n_a})^2$ with ${\bf n_a}$ being some fixed direction, has similar consequences and induces a gap $\hbar \omega_{2k_F} \sim 2 S |K|$. Finally, a magnetic field $B$ gives a gap $\hbar \omega_{2k_F} \sim 2 \mu_S B$, with $\mu_S$ being the magnetic moment of the localized spin.

All these perturbations leave the spectrum at $k=0$ gapless. This is because it stems from symmetry with respect to translations  along the wire (equivalent to rotation of $\boldsymbol{\mathcal{I}}$ around ${\bf h}$), which is not broken by any translationally invariant interaction. This symmetry is broken by finite size effects. Pinning the spins on the wire end(s) by some energy $K$ leads to $k=0$ magnon energy of order $K/N$. A finite system length $L$ defines the smallest available magnon wavevector $k_1=2\pi/L$. It is to be compared to the width of the susceptibility peak $q_w \equiv 2\pi/L_w$, for which we get from Eq.~(5) of the main text the value $L_w=(2\pi/k_F)\sqrt{\mu_F/2\Delta}$. For a wire shorter than $L_w$ there are no low energy magnons---all available ones have energy of order $\hbar \omega \sim S J_{2k_F}$. Then we obtain the  critical temperature of helical order by approximating the magnon energies in Eq.~\eqref{eq:polarization} by a constant, resulting in Eq.~(7) of the main text. We 
list the 
cut-off length $L_w$ for the three realizations in Tab.~SI.

For a wire longer than $L_w$, magnons in the vicinity of $k=0$ get more populated. With a linear spectrum, in the $L\to \infty$ limit, the critical temperature drops logarithmically with $L$. However, since the susceptibility peak is very narrow, we find that Eq.~(7) holds up to exponentially large lengths with a conservative estimate
\be
L \lesssim L_w \exp \left( \frac{L_w}{a} \right),
\label{cutoff}
\tag{S3}
\ee
so that Eq.~(7) holds for any reasonable wire length before fluctuations suppress the helical order.

\begin{table}[h!]
\begin{tabular}{cccc}
\hline\hline
& chain & \quad magnetic wire \qquad & \quad nuclear wire \qquad\\
\hline
$L_w$ & 14 nm & 150 nm & 500 nm\\
\hline\hline
\end{tabular}
\caption{Cut-off length $L_w$.  There are no low energy magnons for wires shorter than $L_w$ leading to the constraint on the wire length $L$ given in Eq.~(\ref{cutoff}).} 
\label{tab:Lw}
\end{table}

\section{Equivalence to spin-orbit coupling}

Let us consider the single-particle Hamiltonian of one-dimensional free electrons under the effect of a rotating magnetic field
\be
H = \frac{\hbar^2\hat{k}^2}{2m} - \mu_F + \mathcal{R}_{{\bf h},2k_Fx} [\boldsymbol{\mathcal{I}}] \cdot \boldsymbol{\sigma}.
\label{eq:Hxg}
\tag{S4}
\ee
We introduce a unitary transformation $\mathcal{H}^\prime = U^\dagger \mathcal{H} U$, corresponding to a  basis change $\psi^\prime(x) = U^\dagger \psi(x)$. Choosing $U=\exp( -i {\bf h}\cdot \boldsymbol{\sigma} k_F x )$, we get that the previous Hamiltonian is equivalent to one with a static magnetic field and a spin-orbit interaction, \cite{Braunecker_Loss_Klin_2009}
\be
\mathcal{H}^\prime = \frac{\hbar^2\hat{k}^2}{2m} - \frac{\hbar^2 \hat{k}}{m l_{\rm so}} {\bf n}_{\rm so} \cdot \boldsymbol{\sigma} + \mu_B {\bf B} \cdot \boldsymbol{\sigma}.
\label{eq:Hxgp}
\tag{S5}
\ee
Here, the spin-orbit length is given by  $ l_{\rm so}=1/2k_F$ ($l_{\rm so} \equiv 1/k_n$).  The spin-orbit vector points along the helical axis ${\bf n}_{\rm so}||{\bf h}$ and the 
magnetic field ${\bf B}$ of strength $\mu_B B=\Delta_m$   points along $\boldsymbol{\mathcal{I}}$.
Importantly, if the spin quantization axis is chosen along ${\bf h}$, the superconducting terms stay invariant, e.g. $\psi^\prime_\uparrow(x) \psi^\prime_\downarrow(x) = \psi_\uparrow(x) \psi_\downarrow(x)$. 

\bibliographystyle{apsrev}

\begin{thebibliography}{100}
\bibitem{footnote1} Subbands can arise from direct hopping inside the wire or  indirect hopping
via the superconductor, or even from coupled Shiba states induced by the magnetic moments \cite{atoms_theory_Princeton,Yazdani_Shiba,Nagaosa}.
\end{thebibliography}

\end{document}